\documentclass[english,pra,showpacs,twocolumn]{revtex4-1}
\usepackage[T1]{fontenc}
\usepackage[latin9]{inputenc}
\usepackage{color}
\usepackage{amsmath}
\usepackage{amssymb}
\usepackage{graphicx}

\makeatletter
\usepackage{txfonts}
\usepackage{dcolumn}
\usepackage{bm}
\usepackage[colorlinks,
            linkcolor=blue,
            anchorcolor=blue,
            citecolor=blue
            ]{hyperref}

\makeatother

\usepackage{babel}
\begin{document}

\title{Generalized Modular Values with Non-Classical Pointer States}

\author{Yusuf Turek$^{1,2}$}
\email{yusufu@csrc.ac.cn}

\author{Taximaiti Yusufu$^{1,2}$ }

\affiliation{$^{1}$School of Physics and Electronic Engineering, Xinjiang Normal
University, Urumqi, Xinjiang 830054, China}

\affiliation{$^{2}$Laboratory of Novel Light Source and Micronano-Optics, Xinjiang
Normal University, Urumqi, Xinjiang 830054, China }
\begin{abstract}
In this study, we investigate the advantages of non-classical pointer
states in the generalized modular value scheme. We consider a typical
von Neumann measurement with a discrete quantum pointer, where the
pointer is a projection operator onto one of the states of the basis
of the pointer Hilbert space. We separately calculate the conditional
probabilities, $Q_{M}$ factors, and signal-to-noise ratios of quadrature
operators of coherent, coherent squeezed, and Schrödinger cat pointer
states and find that the non-classical pointer states can increase
the negativity of the field and precision of measurement compared
with semi-classical states in generalized measurement problems characterized
by the modular value. 
\end{abstract}

\pacs{42.50.Gy, 42.25.Bs,78.20. Ci, 42.25. Kb}
\maketitle

\section{Introduction}

The weak measurement, as a generalized von Neumann quantum measurement
theory, was proposed by Aharonov, Albert, and Vaidman in 1988\cite{Aharonov(1988)}.
In the weak measurement, the coupling between the pointer and the
measured systems is sufficiently weak, but its induced weak value
of the observable on the measured system can be beyond the usual range
of eigenvalues of that observable\cite{qparadox(2005)}. The feature
of weak value is usually referred to as the amplification effect for
weak signals rather than a conventional quantum measurement, and this
amplifying effect occurs when the pre- and post-selection states of
the measured system are almost orthogonal. The successful post-selection
probability tends to decrease to maintain a successful amplification
effect. For more details about weak measurement and weak values, consult
these reviews \cite{Dressel(2014),Nori,Shikano(2010)}.

To date, most weak measurement studies have focused on using the zero-mean
Gaussian state as an initial pointer state. However, recent works
\cite{Wu,Knee(2014)} have showed that the zero-mean Gaussian pointer
state cannot improve the signal-to-noise ratio (SNR)when considering
post-selection probability. A Gaussian beam is classical and it is
natural to inquire about using non-classical pointer states and their
advantages. This issue has been recently addressed \cite{Pang(2014)},
where coherent and coherent squeezed states were utilized as pointers.
The results showed that the post-selected weak measurement improved
the SNR compared with the non-post-selected process if the pointer
state is non-classical rather than classical. The focus of the calculation
was on the assumption that the coupling between measuring device and
measured system is too weak; hence, it is sufficient to consider the
time evolution operator up to its first order. Furthermore, there
have been recent studies giving full-order effects of the unitary
evolution resulting from the von Neumann interaction, but for classical
and semi-classical states\cite{Turek,Nakamura(2012),Turek(2015)}.
Here, we should mention that the standard weak measurement theory
and its induced amplification effect only can be used in weak coupling
strength regions; it is therefore worthwhile to investigate if there
is any new measurement method that is effective for all measurement
strengths.

In 2010, Kedem and Vaidman\cite{Vaidman(2010)} considered the interaction
between a system and pointer qubit, where the system is the weak measurement
conditioned by initial and final states, and the initial state of
the pointer is prepared as $\gamma\vert0\rangle+\beta\vert1\rangle$,
with $\vert\alpha\vert^{2}+\vert\beta\vert^{2}=1$. Although their
scheme can solve the above problem, this kind of one-qubit pointer
state is a limitation for further use of modular values in practical
issues, such as state tomography and non-local measurement. In recent
works, N. Imoto et al.\cite{Ho(2017),Ho(2016)} overcame that limitation
by introducing the generalization modular value scheme; in their model,
the pointer is not a qubit but a qudit. In this multilevel pointer
system, the resulting value is called a generalized modular value.
In this kind of generalized modular value scheme, the pointer projection
operator $\hat{P}$ is not necessarily $\vert1\rangle\langle1\vert$,
but can be any of the eigenstates of $\vert\eta_{\mu}\rangle\langle\eta_{\mu}\vert$
($\mu=0,1,2...,d-1$, where d is the dimension of the qudit pointer).
However, in their work, they only consider semi-classical pointer
states, but the advantages of non-classical pointers in the generalized
modular value scheme is still unclear and requires further study.

In this paper, motivated by the work of N. Imoto et al., we investigate
the extension of the generalized modular value scheme with non-classical
pointer states. We separately consider the coherent, coherent squeezed,
and Schrödinger cat states in the Fock-state basis as pointers and
study the advantages of non-classical pointer states over semi-classical
states in generalized modular value basis measurement problems. We
found that, similar to the standard weak value, the modular value
also has an amplification effect, and to increase the precision of
measurement and negativity of the field process, the non-classical
pointer states have many advantages compared with the semi-classical
states in the generalized modular value scheme.

The rest of the paper is organized as follows. In Section II, we give
the setup for our system and study the relationships between the standard
weak value and modular value. In Section III, we study the advantages
of non-classical pointer states in the generalized post-selected modular
value scheme by investigating the conditional probabilities, negativity,
and SNR of quadrature operator weak measurements of coherent, coherent
squeezed, and Schrödinger cat pointer states. We give the conclusion
to our paper in Section IV. Throughout this paper, we use the unit
$\hbar=1$.

\section{setup and modular value}

For weak measurement, the coupling interaction between the system
and detector is taken to be the standard von Neumann Hamiltonian:
\begin{eqnarray}
H & = & g\delta(t-t_{0})\hat{A}\otimes\hat{P},\label{eq:Hamil}
\end{eqnarray}
where $g$ is a coupling constant and $\hat{P}$ is the conjugate
momentum operator to the position operator $\hat{X}$ of the measuring
device, i.e., $[\hat{X},\hat{P}]=i\hat{I}$. We have taken the interaction
to be impulsive at time $t=t_{0}$ for simplicity. For this kind of
impulsive interaction, the time evolution operator becomes $e^{-ig\hat{A}\otimes\hat{P}}$.
We know that the standard weak measurement is characterized by the
pre- and post-selection of the system state. If we prepare the initial
state $\left|\psi_{i}\right\rangle $ of the system and the pointer
state, after some interaction time $t_{0}$, we post-select a system
state $\left|\psi_{f}\right\rangle $ and obtain information about
a physical quantity $\hat{A}$ from the pointer wave function using
the following weak value: 
\begin{equation}
\langle A\rangle_{w}=\frac{\left\langle \psi_{f}\right|\hat{A}\left|\psi_{i}\right\rangle }{\left\langle \psi_{f}\right|\psi_{i}\rangle}.\label{eq:WV}
\end{equation}
In general, the weak value is a complex number. From Eq. $\left(\ref{eq:WV}\right)$,
we know that when the pre-selected state $\left|\psi_{i}\right\rangle $
and the post-selected state $\left|\psi_{f}\right\rangle $ are almost
orthogonal, the absolute value of the weak value can be arbitrarily
large. This feature leads to weak value amplification, and for most
of the cases, we can use a continuously variable system as a pointer,
such as a Gaussian beam.

However, if we use the standard von Neumann-type Hamiltonian, Eq.
$(\ref{eq:Hamil})$ with a discrete pointer state, the expectation
value of the outcome of such a measurement give the so-called modular
value\cite{Vaidman(2010)}. The modular value for a system observable
$\hat{A}$ is defined as 

\begin{equation}
(\hat{A})_{mod}=\frac{\langle\psi_{f}\vert e^{-ig\hat{A}}\vert\psi_{i}\rangle}{\langle\psi_{f}\vert\psi_{i}\rangle}.\label{eq:MV}
\end{equation}
From the definition of the weak value and modular value given in Eq.
$(\ref{eq:WV})$ and Eq. (\ref{eq:MV}), we know that the modular
value is effective for arbitrarily large coupling strength $g$ and
for discrete pointer states, for which the projection operator is
chosen as $\hat{P}=\vert1\rangle\langle1\vert$, and the initial state
of the qubit pointer is prepared to be $\alpha\vert0\rangle+\beta\vert1\rangle$,
with $\vert\alpha\vert^{2}+\vert\beta\vert^{2}=1$. 

For two commonly used types of system observables $\hat{A}$, such
as $\hat{A}^{2}=\hat{A}$ and $\hat{A}^{2}=\hat{I}$, the relationships
between the modular value and standard weak value can be derived as

\begin{align}
(\hat{A})_{mod} & =\begin{cases}
1-\langle\hat{A}\rangle_{w}+e^{-ig}\langle\hat{A}\rangle_{w}, & \hat{A}^{2}=\hat{A}\\
\cos g-i\langle\hat{A}\rangle_{w}\sin g, & \hat{A}^{2}=\hat{I}
\end{cases}~\label{eq:Rel}
\end{align}
Apparently, for the discrete pointer case, the weak value, $\langle\hat{A}\rangle_{w}$,
can be seen as a result of the weak coupling strength of the modular
value, $(\hat{A})_{m}$.

If we assume that the system is initially prepared to $\vert\psi_{i}\rangle$,
and the $\hat{P}$ is the projection operator $\hat{P}=\vert m\rangle\langle m\vert$
of the pointer whose initial state can be written as 
\begin{equation}
\vert\phi\rangle=\sum_{n}c_{n}\vert n\rangle
\end{equation}
, then after post-selection to state $\vert\psi_{f}\rangle$, the
normalized final state of the pointer is given as 
\begin{equation}
\vert\varphi\rangle=\frac{1}{\delta}\sum_{n}c_{n}(A)_{G}\vert n\rangle\label{eq:fin}
\end{equation}
where $\delta=[1-\vert c_{m}\vert^{2}+\vert c_{m}\vert^{2}\vert\vert(A)_{mod}\vert^{2}]^{\frac{1}{2}}$,
and 
\begin{equation}
(A)_{G}=\frac{\langle\psi_{f}\vert e^{-ig\hat{A}\delta_{nm}}\vert\psi_{i}\rangle}{\langle\psi_{f}\vert\psi_{i}\rangle}\label{eq:genMV}
\end{equation}

is called the generalized modular value.

In this paper, we assume that the operator to be observed is the spin
$x$ component of a spin- $1/2$ particle through the von Neumann
interaction 
\begin{equation}
A=\sigma_{x}=\vert\uparrow_{z}\rangle\langle\downarrow_{z}\vert+\vert\downarrow_{z}\rangle\langle\uparrow_{z}\vert,
\end{equation}
where $\vert\uparrow_{z}\rangle$ and $\langle\downarrow_{z}\vert$
are eigenstates of $\sigma_{z}$ with the corresponding eigenvalues
$1$ and $-1$, respectively. When we select the pre- and post-selected
states as 
\begin{equation}
\vert\psi_{i}\rangle=\cos\theta_{1}\vert\uparrow_{z}\rangle+e^{i\varphi_{1}}\sin\theta_{1}\vert\downarrow_{z}\rangle,\label{eq:Pre}
\end{equation}
and 
\begin{equation}
\vert\psi_{f}\rangle=\vert\uparrow_{z}\rangle,\label{eq:Post}
\end{equation}
respectively, we can obtain the weak value by substituting these states
into 
\begin{equation}
\langle A\rangle_{w}=\langle\sigma_{x}\rangle_{w}=\frac{\langle\psi_{f}\vert A\vert\psi_{i}\rangle}{\langle\psi_{f}\vert\psi_{i}\rangle},
\end{equation}
obtaining 
\begin{equation}
\langle A\rangle_{w}=e^{i\varphi_{1}}\tan\theta_{1}.\label{eq:WV-1}
\end{equation}
where $\theta\in[0,\pi]$ and $\varphi\in[0,2\pi)$. Here, the post-selection
probability is $P_{s}=\cos^{2}\theta_{1}$.\textcolor{black}{{} Throughout
this paper, we use the above pre-selected and post-selected states
and weak value, which are given in Eq.(\ref{eq:Pre},\ref{eq:Post})
and Eq.(\ref{eq:WV-1}) with $g=\frac{\pi}{2}$ and $\varphi_{1}=\frac{\pi}{2}$
for our discussion. }

\section{Modular values with non-classical pointer states}

In this section, we study the general modular values of classical
coherent state and non-classical states, coherent squeezed, and Schrödinger
cat pointer states for arbitrary measurement strength $g$. To show
the advantages of non-classical pointer states in the generalized
modular value scheme:

(1) we check the conditional probabilities of finding photons after
post-selected measurement, which is characterized by generalized modular
values. In our scheme, the conditional probability of finding the
boson numbers $n$ in the field after the post-selected measurement
is given by 
\begin{equation}
p(n)=\vert\langle n\vert\varphi\rangle\vert^{2}=\begin{cases}
\frac{\vert c_{m}\vert^{2}}{1-\vert c_{n}\vert^{2}+\vert c_{n}\vert^{2}\vert\vert(A)_{mod}\vert^{2}} & n\neq m\\
\frac{\vert c_{n}\vert^{2}\vert(A)_{mod}\vert^{2}}{1-\vert c_{n}\vert^{2}+\vert c_{n}\vert^{2}\vert\vert(A)_{mod}\vert^{2}} & m=n
\end{cases}\label{eq:cond prob}
\end{equation}

(2) to investigate the effects of modular values on the non-classicality
of normalized pointer states after the post-selection process, we
check the $Q_{M}$- factor, which is defined as\cite{Awarwal}

\begin{equation}
Q_{M}=\frac{\langle(\triangle n)^{2}\rangle-\langle\hat{n}\rangle}{\langle\hat{n}\rangle}=\frac{\langle a^{\dagger}a^{\dagger}aa\rangle-\langle a^{\dagger}a\rangle^{2}}{\langle a^{\dagger}a\rangle}.\label{eq:MF}
\end{equation}

(3) we discuss the SNR of the quadrature operator $\hat{X}_{\theta}=\frac{1}{\sqrt{2}}(ae^{-i\theta}+a^{\dagger}e^{i\theta})$
with $[\hat{X}_{\theta},\hat{X}_{\theta+\frac{\pi}{2}}]=i$ . The
SNR of the post-selection process is defined as\cite{Awarwal} 
\begin{equation}
SNR_{X}=\frac{\sqrt{NP_{s}}\vert\langle X\rangle_{fi}\vert}{\sqrt{\langle X^{2}\rangle_{f}-\langle X\rangle_{f}^{2}}}.\label{eq:SNR}
\end{equation}
Here, $N$ is the total number of measurements, $P_{s}$ is the probability
of finding the post-selected state for a given pre-selected state,
and $NP_{s}$ is the number of times the system was found in a post-selected
state. Here, $\langle\rangle_{f}$ denotes the expectation value of
measuring the observable under the final state of the pointer.

Next, we separately study the above three quantities for coherent,
coherent squeezed, and Schrödinger cat pointer states for generalized
modular values.

\subsection{Coherent Pointer State}

Here, we assume that the initial state of the pointer is a coherent
state\cite{Scully} of bosons as

\begin{equation}
\vert\alpha\rangle=\sum_{n=0}^{\infty}c_{n}^{\prime}\vert n\rangle,\label{eq:16}
\end{equation}
where $c_{n}^{\prime}=e^{-\frac{1}{2}\vert\alpha\vert^{2}}\frac{\alpha^{n}}{\sqrt{n!}}$
with $\alpha=\gamma e^{i\phi}$. After pre-selection , $\vert\psi_{i}\rangle$
and post-selection, $\vert\psi_{f}\rangle$, the normalized state
of the pointer can be written using Eq. $(\ref{eq:fin})$ by changing
the coefficient $c_{n}$ to $c_{n}^{\prime}$.

\begin{figure}
\includegraphics[width=8cm]{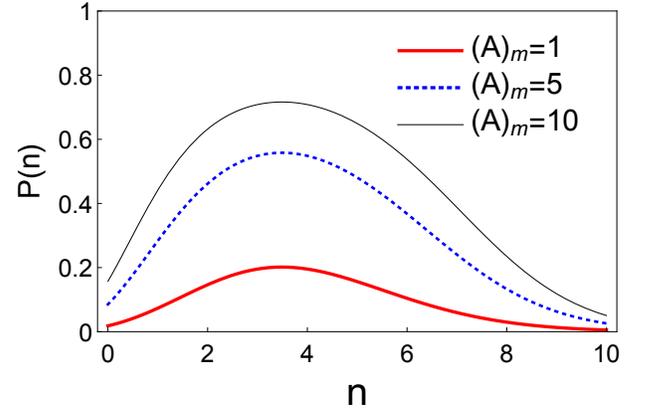}

\caption{\label{fig1}(Color online) For coherent state: the probability $p(n)$
versus $n$ curves with various modular values. Here, $\gamma=2$,
$\phi=0$ $g=\frac{\pi}{2},\text{\ensuremath{\varphi_{1}=\frac{\pi}{2}}}$
and $N=1$. }
\end{figure}

We can obtain the probability of finding the boson number $n$ using
Eq. $(\ref{eq:cond prob})$, and its value with changing modular values
can be seen in Fig.\ref{fig1}. As shown in Fig. \ref{fig1}, the
red curve with $(A)_{m}=1$ represents the probability of finding
the number $n$ without interaction and is a Poisson distribution.
However, as the modular value increases, the probability of finding
the photon in state $n$ increases, demonstrating an amplification
effect of the modular value.

Next, we determine the $Q_{M}$ parameter for a coherent state. Using
the normalized final state of the coherent pointer state 
\begin{equation}
\vert\varphi^{\prime}\rangle=\frac{1}{\delta^{\prime}}\sum_{n}c_{n}^{\prime}(A)_{G}\vert n\rangle,\label{eq:17}
\end{equation}
with $\delta^{\prime}=[1-\vert c_{m}^{\prime}\vert^{2}+\vert c_{m}^{\prime}\vert^{2}\vert\vert(A)_{mod}\vert^{2}]^{\frac{1}{2}}$,
we can obtain 
\begin{equation}
\langle a^{\dagger}a\rangle=\frac{e^{-\vert\alpha\vert^{2}}}{\vert\delta^{\prime}\vert^{2}}\{\vert\alpha\vert^{2}e^{\vert\alpha\vert^{2}}-\frac{\vert\alpha\vert^{2m}m}{m!}[1-\vert(A)_{m}\vert^{2}]\}
\end{equation}
and 
\begin{equation}
\langle a^{\dagger2}a^{2}\rangle=\frac{e^{-\vert\alpha\vert^{2}}}{\vert\delta^{\prime}\vert^{2}}\{\vert\alpha\vert^{4}e^{\vert\alpha\vert^{2}}-\frac{\vert\alpha\vert^{2m}m(m-1)}{m!}(1-\vert(A)_{m}\vert^{2})\}
\end{equation}
respectively.

\begin{figure}
\includegraphics[width=8cm]{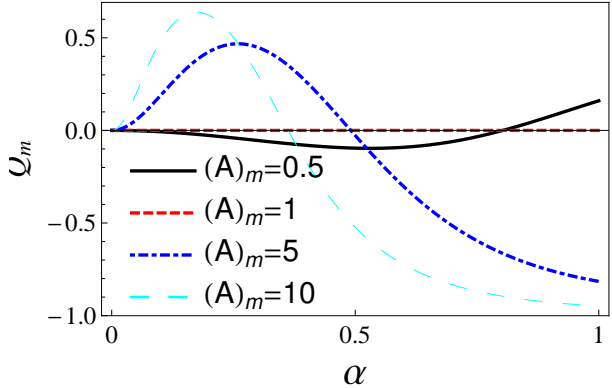}

\caption{\label{fig:fig.2}(Color online) The Mandel $Q_{m}$ parameter as
a function of $\alpha$ for a coherent state. Here, $\gamma=\alpha,\phi=0,$
$m=2$, and other parameters are the same as in Fig. \ref{fig1}. }
\end{figure}

We know that without interaction, $Q_{M}=0$ for a coherent state,
but the negativity of $Q_{M}$ can definitely measure the non-classicality
of the field. From fig. $\ref{fig:fig.2}$, when the modular value
is not equal to one (there is no interaction between the system and
pointer), the factor $Q_{M}$ will always be negative in some region.
Furthermore, if we increase the modular value, its negativity tends
to $Q_{M}=-1$, which corresponds to the Fock state with increasing
coherent state parameter $\alpha$. The coherent state is a typical
semi-classical field, but as seen in Fig. $(\ref{fig:fig.2})$, the
generalized post-selected measurement can change its field characteristics
more dramatically with increasing modular value.

\begin{widetext}
\begin{figure}
\includegraphics{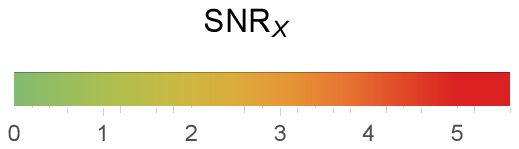}

\centering

\includegraphics[width=5.3cm]{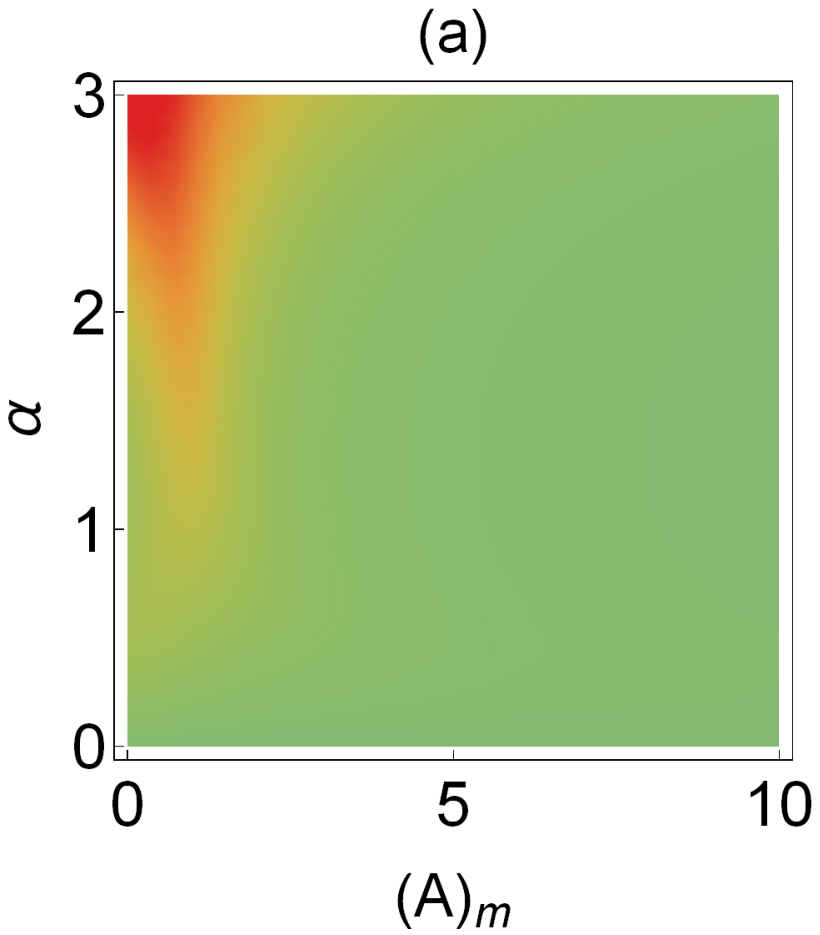}\includegraphics[width=5.3cm]{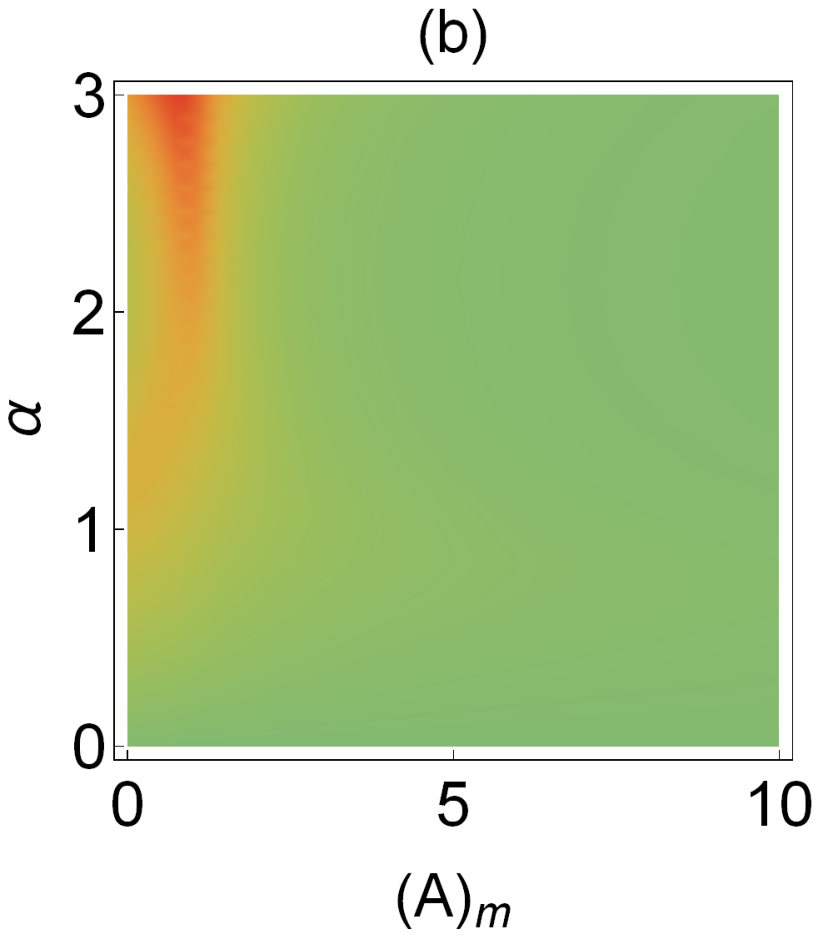}\includegraphics[width=5.3cm]{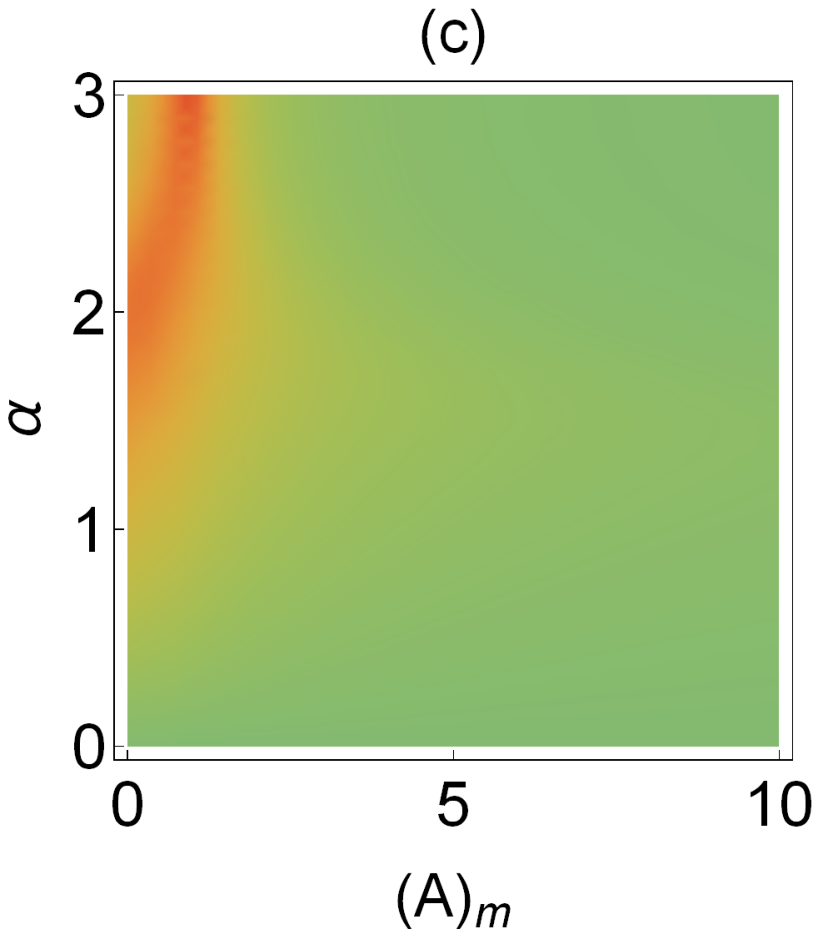}

\caption{\label{fig.3}(Color online) SNR of quadrature operator $X$ for Coherent
pointer states as a function of the modular value $(A)_{m}$ and coherent
state parameter $\alpha$. Here, (a) $m=2$, (b) $m=5$, and (c) $m=10$,
respectively. $\theta=0$(corresponding to the $x$ direction) and
the other parameters are the same as that in Fig. \ref{fig1}.}
\end{figure}
\end{widetext}

To investigate the advantages of modular value in precision measurement,
we check the SNR of quadrature operator $X_{\theta}$ with the considered
the post-selection probability. To do this, we first calculate the
expectation value of $X_{\theta}$ and $X_{\theta}^{2}$ under the
normalized final state, Eq. (\ref{eq:17}), and the results are given
as 
\begin{equation}
\langle X_{\theta}\rangle=\frac{\sqrt{2}e^{-\vert\alpha\vert^{2}}}{\vert\delta^{\prime}\vert^{2}}\Re[\alpha^{\ast}\{((A)_{m}-1)\frac{\vert\alpha\vert^{2m}}{m!}+((A)_{m}^{\ast}-1)\frac{\vert\alpha\vert^{2(m-1)}}{(m-1)!}\}e^{i\theta}+e^{\vert\alpha\vert^{2}}]
\end{equation}
and 
\begin{align}
\langle X_{\theta}^{2}\rangle & =\frac{1}{\vert\delta^{\prime}\vert^{2}}\{\vert\alpha\vert^{2}+\vert c_{m}\vert^{2}(\vert(A)_{m}\vert^{2}-1)m\}+\frac{1}{2}\nonumber \\
 & +\frac{1}{\vert\delta^{\prime}\vert^{2}}\Re\{\alpha^{\ast2}e^{-\vert\alpha\vert^{2}}e^{2i\theta}[e^{\vert\alpha\vert^{2}}+((A)_{m}-1)\frac{\vert\alpha\vert^{2m}}{m!}+((A)_{m}^{\ast}-1)\frac{\vert\alpha\vert^{2(m-2)}}{(m-2)!}]\}
\end{align}
respectively. We also plot the analytical result as a function of
modular value and coherent state parameter $\gamma=\alpha$, and the
result is shown in Fig. $\ref{fig.3}$. As indicated in Fig. $\ref{fig.3}$,
the SNR of $X_{\theta=0}$ increases when the coherent state parameter
$\alpha>1$ has a small modular value. However, the SNR does not increase
significantly with an increase in the modular value.

\subsection{Coherent Squeezed State}

The coherent squeezed state is a typical quantum state. It has many
applications in optical communication, optical measurement, and gravitational
wave detection \cite{Milburn}. Here, we assume that the initial state
of the pointer is a coherent squeezed state \cite{Scully}; in the
Fock-state basis, its definition can be written as

\begin{align}
\vert\alpha,\xi\rangle & =\sum_{n=0}^{\infty}\beta_{n}\vert n\rangle,
\end{align}
where $\gamma=\alpha\cosh r+\alpha^{\ast}e^{i\theta}\sinh r$, and
\begin{align}
\beta_{n} & =\frac{1}{\sqrt{\cosh r}}\exp[-\frac{1}{2}\vert\alpha\vert^{2}-\frac{1}{2}\alpha^{\ast2}e^{i\theta}\tanh r]\times\nonumber \\
 & \frac{[\frac{1}{2}e^{i\theta}\tanh r]^{\frac{n}{2}}}{\sqrt{n!}}H_{n}[\gamma(e^{i\theta}\sinh(2r))^{-\frac{1}{2}}
\end{align}
The normalized function of the coherent squeezed state after post-selection
is given as

\begin{align}
\left|\psi\right\rangle  & =\frac{1}{\eta}\{\beta_{m}[(A)_{m}-1]\vert m\rangle+\vert\alpha,\xi\rangle\}\label{eq:24}
\end{align}
where 
\begin{equation}
\eta=[1-\vert\beta_{m}\vert^{2}+\vert\beta_{m}\vert^{2}\vert\vert(A)_{mod}\vert^{2}]^{\frac{1}{2}}
\end{equation}

\begin{figure}
\includegraphics{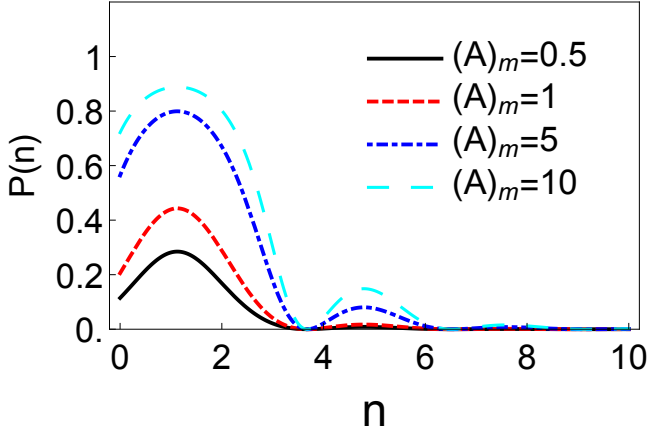}

\caption{\label{fig.4}(Color online) The photon probability of squeezed state
vs. modular value. Here, $\theta=0$, $r=0.5$, $\alpha=1$, and other
parameters are the same as in Fig. \ref{fig1}. }
\end{figure}

As a coherent state case, we first investigate the effect of modular
values on the probability of finding $n$ bosons after post-selection
generalized projection measurement; the analytical result can be calculated
using Eq. $\ref{eq:cond prob}$ by changing $c_{n}$ to $\beta_{n}$.
As shown in Fig. $\ref{fig.4}$, compared with the no interaction
case, $(A)_{m}=1$, the probability of finding $n$ photons increases
with increasing modular value, and this process can also be seen as
a result of the amplification effect of the modular value.

As a coherent state case, to study the effect of the modular value
on the field properties, we calculate the $Q_{M}$ parameter (see
Eq.$\ref{eq:MF}$) for the coherent squeezed state and discuss the
analytical results. Using the normalized final state of the coherent
squeezed pointer, Eq. (\ref{eq:24}), we obtain 
\begin{equation}
\langle a^{\dagger}a\rangle=\vert\alpha\vert^{2}+\sinh^{2}r-\vert\beta_{m}\vert^{2}m(1-\vert(A)_{m}\vert^{2})
\end{equation}
and 
\begin{align}
\langle a^{\dagger2}a^{2}\rangle & =\vert\alpha\cosh r-\alpha^{\ast}e^{i\theta}\sinh r\vert^{2}+2\sinh^{2}r\cosh^{2}r\nonumber \\
 & +(\vert\alpha\vert^{2}+\sinh^{2}r)(1+\vert\alpha\vert^{2}+\sinh^{2}r)\nonumber \\
 & -\vert\beta_{m}\vert^{2}(1-\vert(A)_{m}\vert^{2})m(m-1),
\end{align}
respectively.

\begin{figure}
\includegraphics[width=4cm]{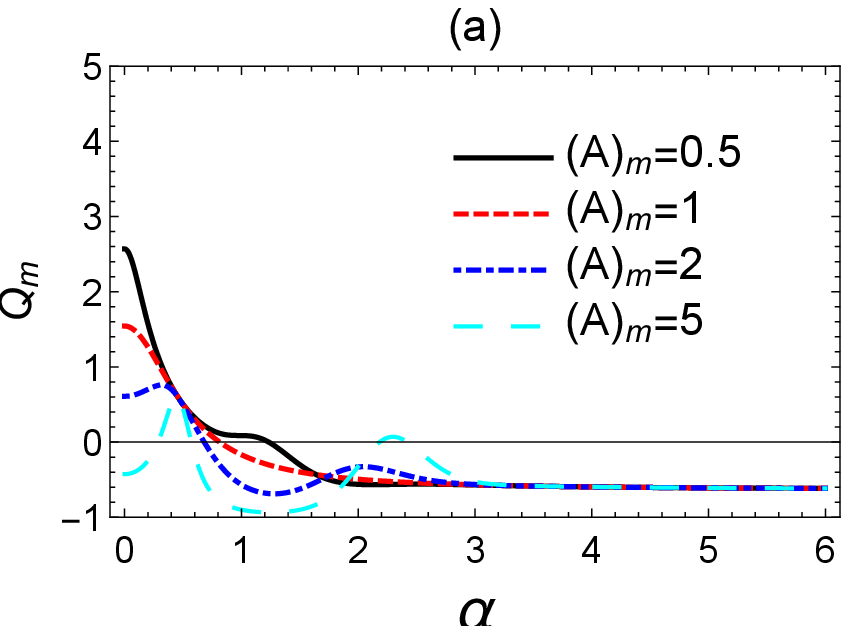} \includegraphics[width=4cm]{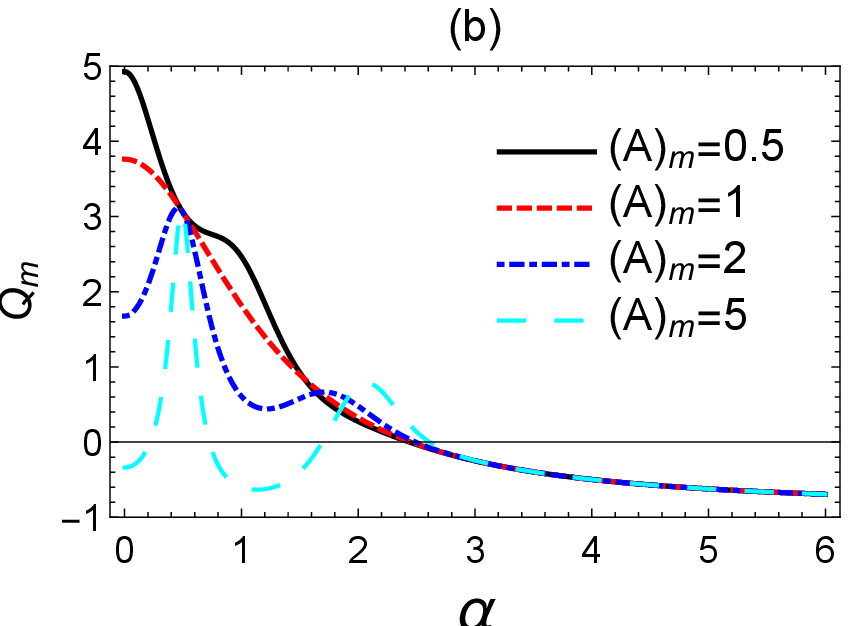}

\caption{\label{fig.5}(Color online) The Mandel $Q_{m}$ factor squeezed coherent
state as a function of coherent parameter $\alpha$ for various modular
values $(A)_{m}$. Here, $m=2$ and (a) $r=0.5$, (b) $r=1$. Other
parameters are the same as in Fig. \ref{fig1}. }
\end{figure}

As shown in Fig. $\ref{fig.5}$, compared with the no interaction
case, $(A)_{m}=1$, the $Q_{M}$ factor for the coherent squeezed
state changes more and its negativity also increases with increasing
modular value. Furthermore, comparing Fig. $\ref{fig.5}$(a) and (b),
it is clear that for the same $\alpha$ and modular value $(A)_{m}$,
the negativity of the coherent squeezed state is more significant
for small squeezing parameter $r$.

For the coherent squeezed state, we also calculate the SNR of quadrature
$X_{\theta}$. The expectation values of $X_{\theta}$ and $X_{\theta}^{2}$
under the normalized final coherent squeezed pointer state, Eq. (\ref{eq:24})
can be calculated as 
\begin{align}
\langle X_{\theta}\rangle & =\frac{\sqrt{2}}{\vert\eta\vert^{2}}\times\nonumber \\
 & \Re\{[\alpha^{\ast}\!\!+\!\!((A)_{m}-1)c_{m+1}^{\ast}c_{m}\sqrt{m+1}+\!\!((A)_{m}^{\ast}-1)c_{m}^{\ast}c_{m-1}\sqrt{m}]e^{i\theta}\},
\end{align}
and 
\begin{align}
\langle X_{\theta}\rangle^{2} & =\frac{1}{\vert\eta\vert^{2}}\{\vert\alpha\vert^{2}+\sinh^{2}r+\vert c_{m}\vert^{2}(\vert(A)_{m}\vert^{2}-1)m\}+\frac{1}{2}\nonumber \\
 & +\frac{1}{\vert\eta\vert^{2}}\Re[((A)_{m}^{\ast}-1)\beta_{m}^{\ast}\beta_{m-2}\sqrt{m(m-1)}e^{2i\theta}]\nonumber \\
 & +\frac{1}{\vert\eta\vert^{2}}\Re\{[\alpha^{\ast2}-e^{-i\phi}\sinh r\cosh r\nonumber \\
 & +((A)_{m}-1)\beta_{m+2}^{\ast}\beta_{m}\sqrt{(m+1)(m+2)}]e^{2i\theta}\}
\end{align}
respectively.

\begin{figure}
\includegraphics[width=4.5cm]{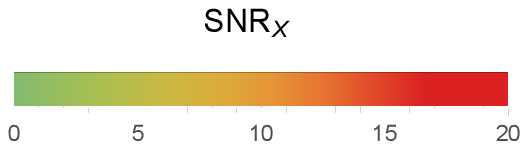}

\includegraphics[width=4cm]{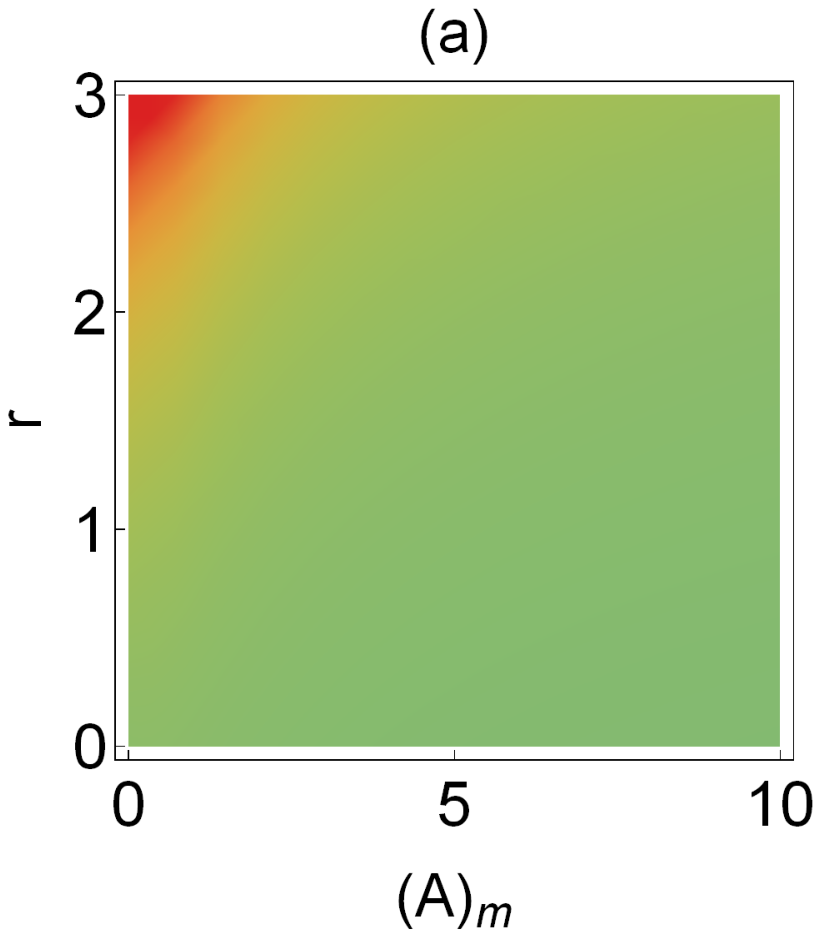}\includegraphics[width=4cm]{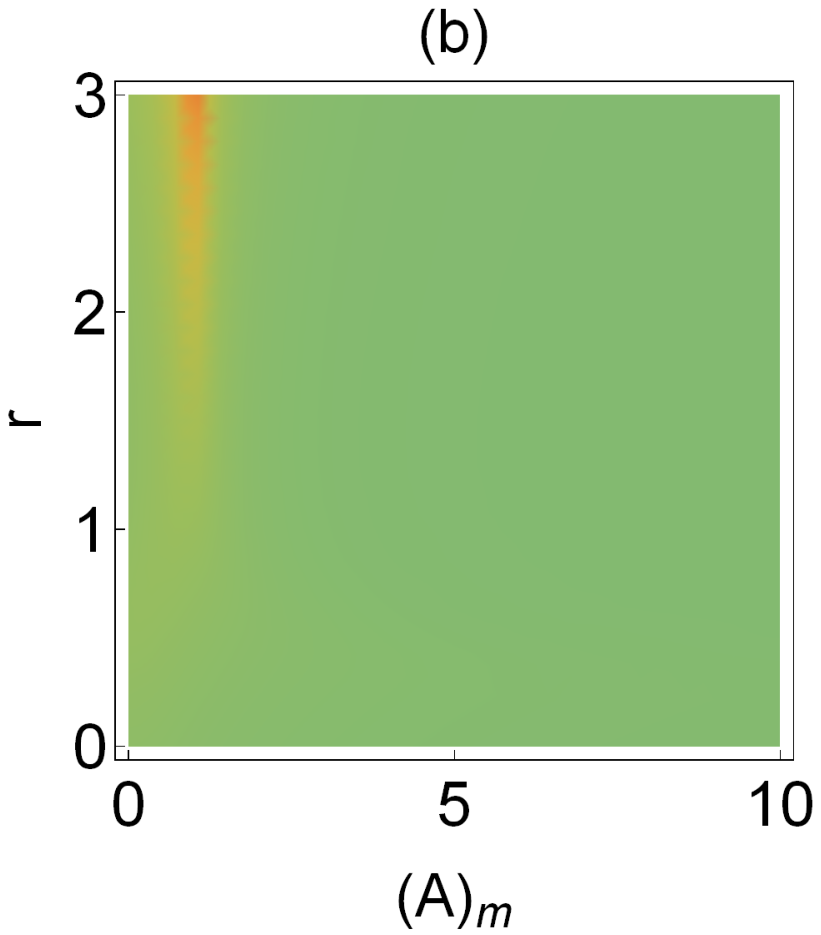}

\caption{\label{fig.6}(Color online) The SNR of $X_{\theta=0}$ for the coherent
squeezed state as a function of modular value and squeezing parameter
$r$: (a) m = 2 and (b) m =5. Here, we take $\alpha=0.5$, and other
parameters are the same as in Fig. \ref{fig1}. }
\end{figure}

The SNR as a function of modular value and the analytical result is
shown in Fig. $\ref{fig.6}$. The $SNR$ increases dramatically with
increasing modular value for definite boson numbers $m$.

\subsection{Schrödinger cat state}

The Schrödinger cat state, another typical quantum state, is a superposition
of two coherent correlated states moving in opposite directions. Generally,
there are two kinds of Schrödinger cat states\cite{Dodonov}; even
and odd Schrödinger cat states. Therefore, we consider the general
Schrödinger cat state as an initial pointer state in the Fock-state
basis to examine the advantages of the non-classical pointer state
in the generalized modular value scheme. The normalized even Schrödinger
cat state\cite{Awarwal} can be written as

\begin{align}
\vert\varphi\rangle & =\mathcal{N}^{-\frac{1}{2}}(\vert\alpha\rangle+e^{i\varphi}\vert-\alpha\rangle)\\
 & =\mathcal{N}^{-\frac{1}{2}}\sum_{n=0}^{\infty}c_{n}^{\prime}\vert n\rangle
\end{align}
where $\mathcal{N=\mathrm{2+2e^{-2\vert\alpha\vert^{2}}\cos\varphi}},$
and $c_{n}^{\prime\prime}=e^{-\frac{\vert\alpha\vert^{2}}{2}}\frac{\alpha^{n}}{n!}(1+e^{i\varphi}(-1)^{n})$.
The normalized state after the post-selected measurement is 
\begin{equation}
\vert\varphi\rangle=w^{-1}\sum_{n}c_{n}^{\prime\prime}(A)_{G}\vert n\rangle,\label{eq:18}
\end{equation}
with 
\begin{equation}
w=[2+2e^{-2\vert\alpha\vert^{2}}\cos\varphi+\vert c_{n}^{\prime\prime}\vert^{2}(\vert(A)_{m}\vert^{2}-1)]^{\frac{1}{2}}.
\end{equation}

\begin{figure}
\includegraphics[width=4cm]{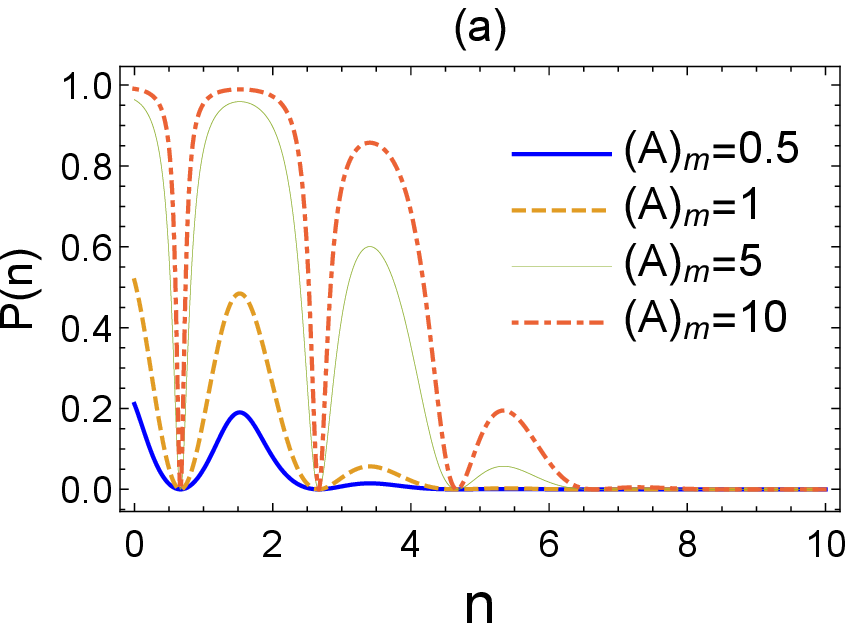}\includegraphics[width=4cm]{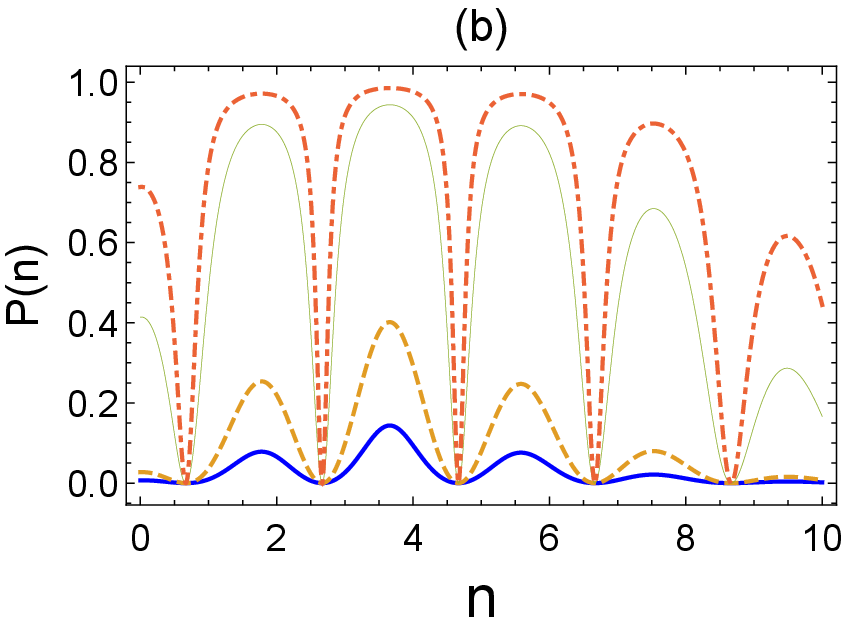}

\caption{\label{fig.7}(Color online) The conditional probability of the Schrödinger
cat state as a function of photon number $n$ with various modular
values $(A)_{m}$. Here, $\varphi=\frac{\pi}{3},$ and (a)$\alpha=1$,
(b) $\alpha=2$. Other parameters are the same as in Fig. \ref{fig1}. }
\end{figure}

The conditional probability of finding $n$ boson numbers of the Schrödinger
cat state after generalized modular value measurement can be calculated
using Eq. $(\ref{eq:cond prob})$ by changing $c_{n}$ to $c_{n}^{\prime\prime}$,
and the analytical results are given in Fig. $\ref{fig.7}$. Compared
with the no interaction case ($(A)_{m}=1$), the conditional probability
increases dramatically with increasing modular value. Furthermore,
by comparing Fig. $\ref{fig.7}$ $(a)$ and $(b)$, it is clear that
for definite modular values, if we increase the coherent state parameter
$\alpha$, the conditional probability also increases dramatically
for all photon numbers and its shape resembles a periodic function.

\begin{figure}
\includegraphics{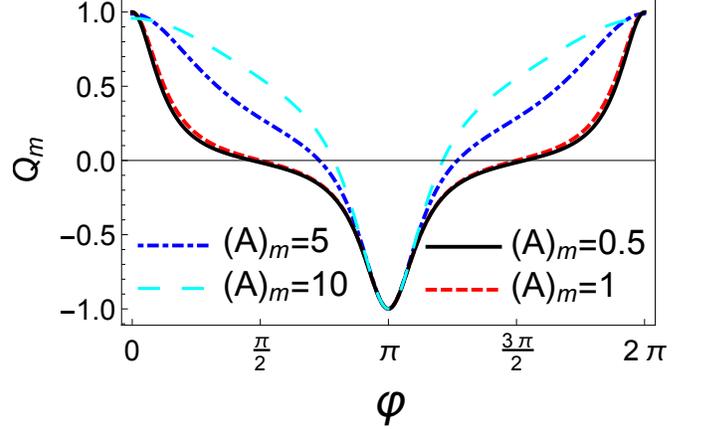}

\caption{\label{fig.8}(Color online) The Mandel $Q_{m}$parameter as a function
of $\varphi$ for the Schrodinger cat state. Here, $m=2$, $\alpha=0.2$,
and other parameters are the same as in Fig. \ref{fig1}. }
\end{figure}

Similar to the coherent state and coherent squeezed state, we investigate
the effect of modular values on the field property of the Schrödinger
cat state by calculating the $Q_{m}$ factor, and the analytical result
is shown in Fig. $\ref{fig.8}$ as a function of angle $\varphi$
for various definite modular values. From Fig. \ref{fig.8}, we can
see that compared with the no interaction case (blue dotted curve),
the negativity curves become more narrow with increasing modular value,
and this also can be seen as a result of the amplification effect
of the modular value. Thus, we can conclude that the modular value
can increase the non-classicality of states.

Finally, we study the SNR of quadrature operator $X_{\theta}$ of
the Schrödinger cat state using the definition of SNR, i.e., Eq. $(\ref{eq:SNR})$.
For the Schrödinger cat state, the expectation value of $X_{\theta}$
and its square can be given as

\begin{align}
\langle\hat{X}_{\theta}\rangle & =\langle\psi\vert\hat{X}_{\theta}\vert\psi\rangle\nonumber \\
 & =\frac{2}{\sqrt{2}\delta^{2}}\Re[\{((A)_{m}-1)c_{m+1}^{\ast}c_{m}\sqrt{m+1}+((A)_{m}^{\ast}-1)c_{m}^{\ast}c_{m-1}\sqrt{m}\nonumber \\
 & +\alpha^{\ast}(2+2i\sin\varphi e^{-2\vert\alpha\vert^{2}})\}e^{i\theta}],
\end{align}
and 
\begin{align}
\langle\hat{X}^{2}\rangle & =\langle\psi\vert\hat{X}^{2}\vert\psi\rangle\nonumber \\
 & =+\frac{1}{\delta^{2}}\Re\{[2\alpha^{2}(1+\cos\varphi e^{-2\vert\alpha\vert^{2}})+((A)_{m}-1)c_{m+2}^{\ast}c_{m}\sqrt{(m+1)(m+2)}\nonumber \\
 & +((A)_{m}^{\ast}-1)c_{m}^{\ast}c_{m-2}\sqrt{m(m-1)}]e^{2i\theta}\},
\end{align}
respectively.

To study the advantages of the modular value for precision measurement
with the Schrödinger cat pointer state, we plotted the SNR of the
quadrature operator $X_{\theta}$ as a function of modular value $(A)_{m}$
and state parameter $\alpha$; the results are given in Fig.$\ref{fig.9}$.
The results apparently show that the SNR increases for all The modular
values with a large parameter $\alpha$; this result is different
from those for the coherent and coherent squeezed pointer cases (see
Fig. \ref{fig.3} and Fig. \ref{fig.6}).

\begin{widetext}
\begin{widetext}
\begin{figure}
\includegraphics{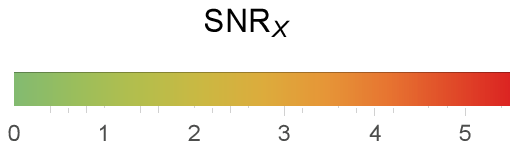}

\includegraphics[width=5.3cm]{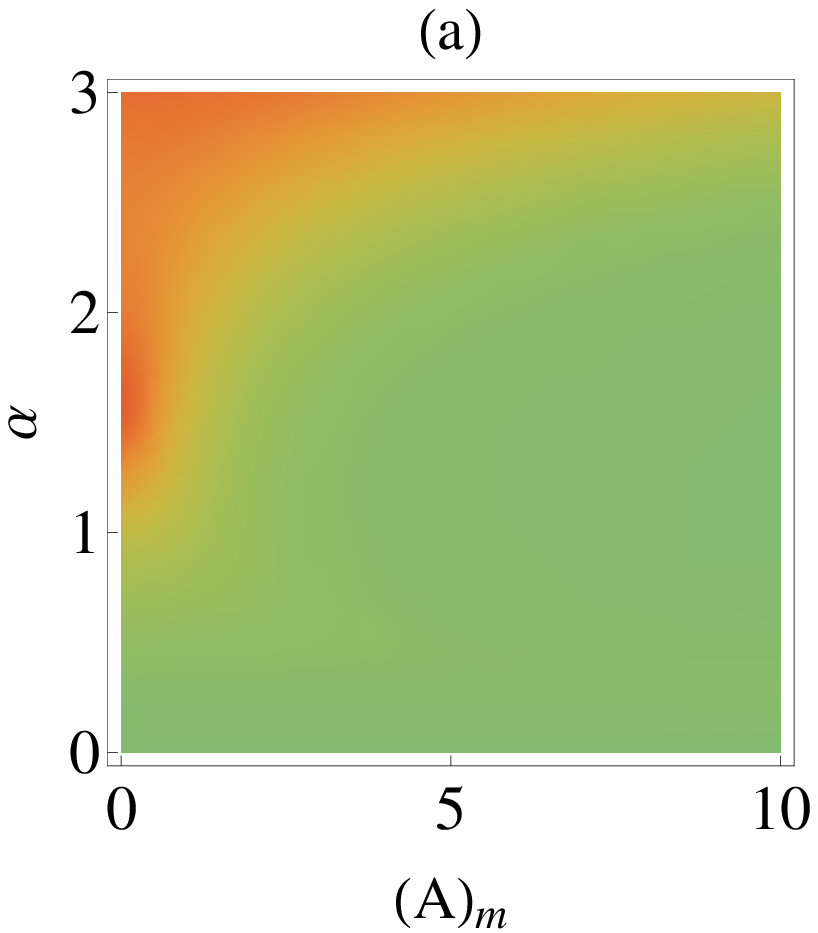}\includegraphics[width=5.3cm]{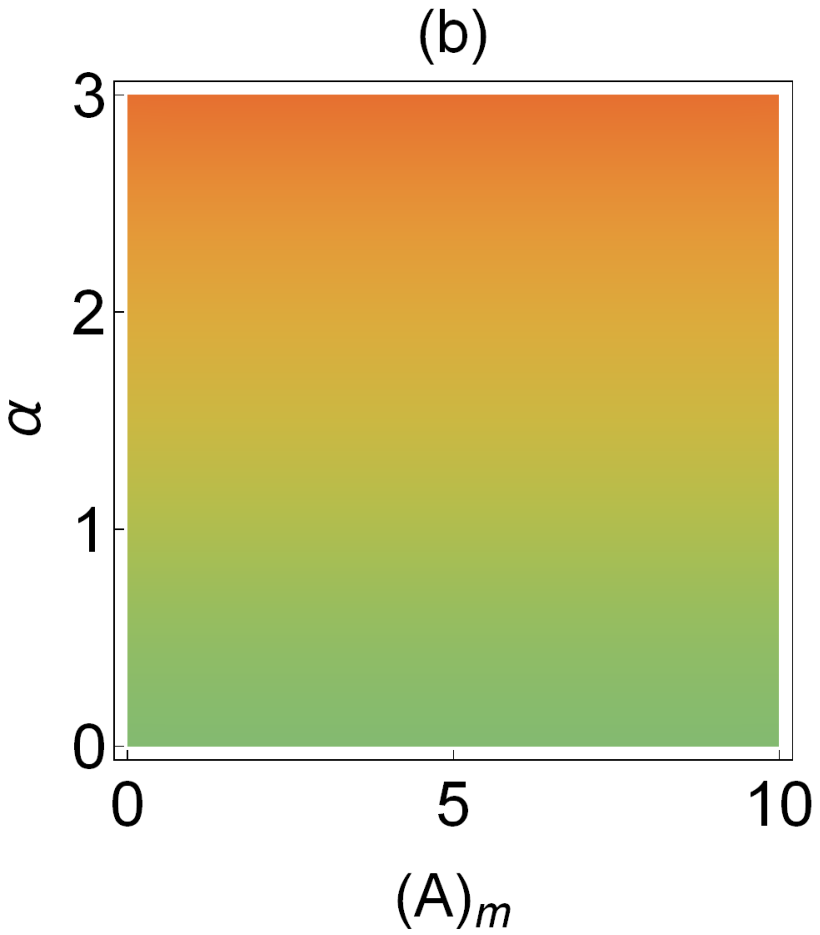}\includegraphics[width=5.3cm]{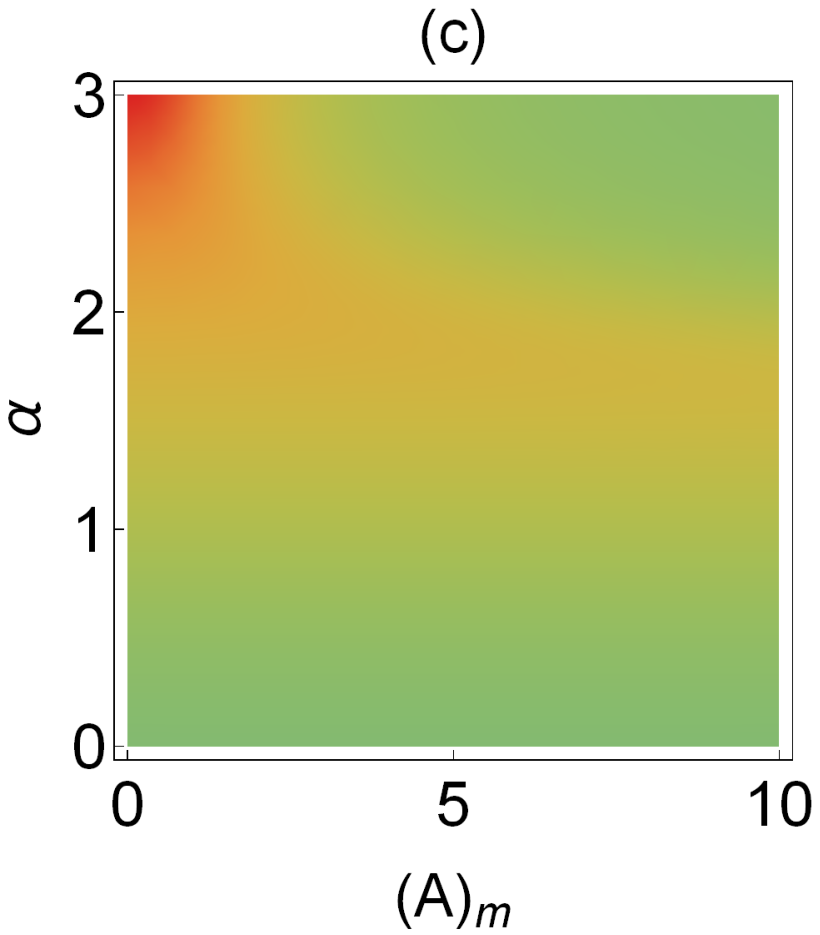}

\caption{\label{fig.9}(Color online) SNR in the x-direction for Schrodinger
pointer states with the modular value with respect to the parameter
$\alpha$ (a) m = 2, (b) m = 5, and (c) m = 10. We use $\phi=0$ in
equation (22) in all figures and $\theta=0$,$\varphi=0$, and other
parameters are the same as in Fig. \ref{fig1} }
\end{figure}
\end{widetext}

\end{widetext}

\section{Conclusion}

We have studied the generalized modular value scheme with semi-classical
and non-classical pointer states. We reported the relationship between
the standard weak value and modular value. By calculating the conditional
probability, $Q_{M}$ factor, and SNR of quadrature operator $X_{\theta}$
for some typical pointer states, such as the coherent state, coherent
squeezed state, and Schrödinger cat state, we show the advantages
of a non-classical pointer in the generalized modular value scheme
compared with the semi-classical one (coherent state). We found that
the conditional probabilities of finding $n$ boson numbers after
post-selected measurement increased with increasing modular values.
We also found that the modular values can change the negativity of
the quantum field, and the non-classical field becomes more non-classical.
In particular, after post-selected measurement with the modular value,
the semi-classical coherent field changed to a non-classical field
in some regions, and its non-classicality increased with increasing
modular value. With respect to the SNR, we found that the SNR of quadrature
operator $X_{\theta}$ of the coherent squeezed state and Schrödinger
cat pointer state increased dramatically with increasing modular value
compared with the coherent state pointer case. Our paper is a generalization
of the original one-qubit pointer state modular value scheme and can
be used in future works related to the application of the modular
value theory with high-dimensional non-classical pointer schemes.
\begin{acknowledgments}
The author would like to thank the hospitality of Prof. Sun to support
a platform to accomplish this project. This work was supported the
National Natural Science Fund Foundation of China (Grant No. 11664041),
and the Doctoral Scientific Research Foundation of Xinjiang Normal
University (Grant No. XNNUBS1807). 
\end{acknowledgments}

\end{document}